\documentclass{article}
\usepackage{a4}
\usepackage{times}
\usepackage{amssymb}
\usepackage{graphicx}
\begin{document}
\title{The 115 GeV signal from nuclear physics.}
\author{Alejandro Rivero \thanks{EUPT (Universidad de Zaragoza), 44003 Teruel, Spain} }
\maketitle

\begin{abstract}
According standard models of nuclei, steepest variations of binding
energy at the drip lines should happen for nuclear masses of
45, 68, 92, 115, 175 and 246 GeV. We explore the coincidence of
these masses with another ones from well known HEP research and we
wonder how much, of it, is an experimental measurement coming from low
energy research. 
\end{abstract}

\section{Signals to be pursued}

At the time of the closure of LEP2, a deviation from background at 115 GeV
was estimated to be worth of further investigation, and it still remains
as the only one worth which the researches felt that more data was needed. 
Besides it, a similar phenomena happened in 1984, when UA1 \cite{UA1, nature}
announced a 40-45 GeV deviation suspected of being a top quark -discarded
after further data collection and modeling- and around 1998, when
L3 \cite{L3a,L3b} got almost four sigma deviation at 68-70 GeV, suspected
of being a charged scalar -discarded after consideration of new background data
from the collected database-.

To these signals, we add the well known ones in the massive side of the
spectra, namely the quark top about 175 GeV and the W and Z0 particles
at 81 and 92 GeV respectively. As explained in the corresponding section,
we do not expect W to have a role similar to the rest of signals. 

The quark top being a confined particle, one could expect to pursue instead
the family of top based mesons, also around the same mass.

Finally, a natural scale in the GeV range is the one of electroweak vacuum
expected value, 246 GeV. We add it to the list in the suspicion that some
models could favour some particle near it. 

We can not assess how biased our selection is, specially for the three values in
the first paragraph. In their age, they generated more attention than other
deviations, including --in two cases-- editorial comments from Nature. But
an exhaustive list of deviations from background for all the 1-300 GeV range
is lacking, partly because of the long time span, partly because model
independent searches are very difficult to implement. In the Tevatron
such search has been attempted \cite{tevatron}, pointing to some high energy
events no far from the above suggested 246 GeV level, but not being too concrete
about lower levels.
Thus, by now, exhaustiveness of this list is argued only from the memory of the
author, who do not remember any other signal so interesting as the three first
ones. Given the extraordinary consistence of the data, memory repression can
not be discarded, and I will thank any information about forgotten signals.

\section{Drip lines matter}

For a nucleus of atomic number $A=N+Z$ in the beta stability line, we can consider the corresponding 
nuclei $(Z-k,N)$ in the proton drip line and $(Z,N-k')$ in the neutron drip line, with respective
masses $A_p, A_n$. The main mass models in the market (eg, from \cite{uno}) predict a very
small difference $A_p-A_n$, which even becomes zero in isolated points under the action of
microscopical corrections.  

We have studied this difference for the classical Weizs\"acker formula\cite{bethe}, 
$$E_b=a_1 A - a_2 A^{2/3} - a_3 {Z^2 \over A^{1/3}} - a_4 {(A-2Z)^2 \over A}$$

An analytical -even if very large- expression can be given if instead of taking $A$ as the
independent variable, we fix the mass $A_0$ in the drip lines. Then solving the second degree
equation in the proton drip line
$$M[Z, A_0] - M[Z - 1, A_0 - 1] - m_z=0$$
and the third degree one in the stability line (we take $m_p-m_n\sim 0$ but it is not necessary)
$$
Z={2 a_4 A \over a_3 A^{2/3} + 4 a_4}
$$
we can get the corresponding mass A and proton and neutron numbers $Z, N (=A-Z)$ of the
stable nucleus. We compare this neutron number with the one got from the
neutron drip line equation
$$M[A_0 - N_0, A_0] - M[A - N_0, A_0 - 1] - m_n =0.$$
The difference $d(A_0)=N-N_0$ results a very convenient function to input in
a numerical-analytical program such as, for instance, {\it Mathematica}, because we can plot 
dependences with any of the four free parameters of the model, as well as mixed plots 
$d(A_0,a_i)$ or $d(a_i,a_j)$. It is specially relevant to check the dependence in $a_3$,
because it has a natural minimum for the zero value, but it is not uniformly increasing;
there is a second minimum in the $\sim 1$ MeV  area, but this one has also a
dependence on $A_0$ so we can no expect it to coincide exactly with the usual
value $a_3 = 0.711$ MeV. Still, this minimum can be interpreted as the cause of our 
equidistancy.

For the usual values
$$a_1 = 15.75 \mbox{MeV}, a_2 = 17.8 \mbox{MeV}, a_3 = 0.711 \mbox{MeV}, a_4 = 23.0 \mbox{MeV},$$
it can be seen
that $d(A_0)$ gets the higher value for $A_0\sim 300$; it is only -0.815 
when proton and neutron masses are equal, and this maximum discrepancy moves by only
about two units when proton and neutron masses are given different value, so for
simplicity one can find convenient to keep with $m_p-m_n\sim 0$ as we do here. 

The function is not linear, so for mid-range masses the difference is appreciably smaller. Generically we
can say that the equidistance property $k=k'$ with different proton and neutrons masses holds 
within a 2\%.
 
As we have said, it can be noticed in most models of nuclear masses, and
our function $d(A_0)$, or alternatively any measure of the discrepancy between $k$ and $k'$, is an
 interesting parameter to consider when studying the properties of a mass model.

An explanation of this property should be that really some important mass
dependent effect is enhanced at the drip lines, so this effect forces mass formulae to
adjust themselves to fit. The effect does not need to happen uniformly in all the 
mass range, it is enough to force the coincidence in isolated points, perhaps the ones
having strong microscopical corrections, as noted above.

\section{Magic numbers}

When we consider the points of steepest change in binding energy -the
traditionally "magic numbers"-, a new relationship between driplines adds
to the one revealed in the previous section. A nucleus having at the proton drip 
line a magic (or semimagic) Z number will correspond to a nucleus of the
same mass at the neutron drip line and a corresponding N magic or semimagic
number.

The correspondence follows this table: 

 \begin{tabular}{l || c | c| c  |c |c |c}
 At neutron dripline, N= & 28-30 & 50 & (64) & 82 & 126 & 184 \\
 \hline
 At proton dripline, Z= & 28 & (40)  & 50 & (58) & 82 & 114
 \end{tabular}

Let us to stress that the correspondence has being done in terms
of mass values at drip lines. Of course, due to the equidistance rule
of previous section, these magicities will meet near the beta
stability line, then causing very strong double magic nucleus.
Particularly interesting are Z=40, N=50 and Z=58, N=82, points
that are too evident -even excesively- when studying beta decay of
unstable nuclei. 

To resume, we have two sources of evidence of some effect in the
nuclear driplines: on one side, nuclear models adjust themselves
in a way such that they are equidistant to stability. On other
side, the position of the drip lines stablishs a correspondence 
between magic (or semimagic)
numbers for Z and N.

Now, let me to ennounce the main result of the paper: the atomic
mass values relevant for this correspondence at the drip lines are,

\begin{tabular}{l || c | c| c |c |c |c}
 Atomic mass (in GeV) & $\sim 44$ & 68  & 91.2 & 115 & 175 & 246 \\
 Atomic mass (in u.) & $\sim 47$ & 73& 98& 123 &187 & 264 
\end{tabular}

There is a variation about a few percent depending of the mass model chosen
to predict the drip lines, but it is mostly negligible.

Thus we can suspect that the effect at drip lines is related to mass values
of elementary particles.

\section{Deviations in semi empirical models}

\begin{figure}[!htb]
    \centering
    \includegraphics[width=6cm]{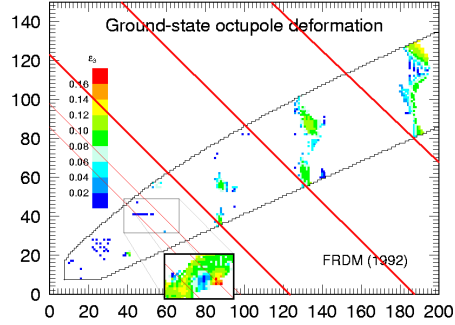}
    \caption{figure 10 of \cite{droplet}, plus an inset -showing local error- of figure 20. We translate
    between GeV and atomic mass units via the conversion constant $1\mbox{u.}=0.9315 \mbox{GeV}$.
    The only addition to the original plot are the diagonal 
    isobars, at $M_W$, $M_Z$, 115 Gev, 
    $M_t$ and 246 Gev.}
    \label{e3}
\end{figure}

Historically, we found the above relationship while examining small
corrections in mass models. 
The main clue came from 1992 FRDM. It shows error at W and Z, but the fit at
other energies is right. Studying the model, we learn that the additional
precision is got from a series of microscopic corrections and shape 
corrections. 
Figure \ref{e3} shows all nuclei where a extra correction $\epsilon_3$ is
applied. We have taken directly the plot from \cite{droplet}, only adding
the diagonal isobars. Neutron dripline is exactly the one drawn by the
original authors ten years ago. This clue pointed us to study the drip
lines instead of the filling procedure (which we did in \cite{models})

\begin{figure}
    \centering
   \includegraphics[width=9cm]{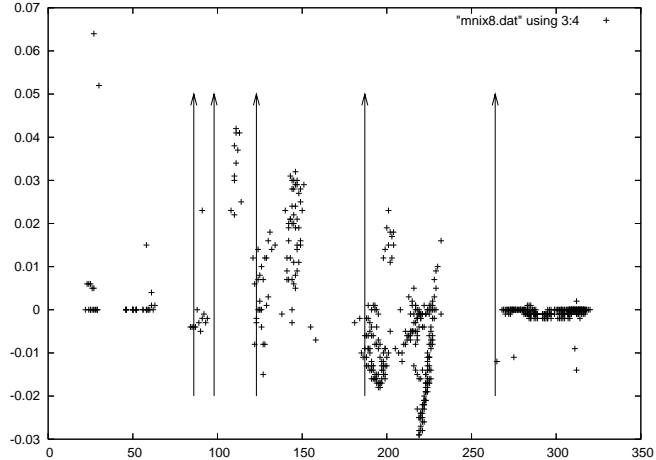}
    \caption{values of $\epsilon_6$ discarded in the FRDM when using
    instead the deformation parameter $\epsilon_3$. Note the qualitative
    difference in the signal of electroweak vacuum.}
    \label{e6sym}
\end{figure}

We were surprised with the apparition of the 246 GeV scale, which, as 
said above, is simply the vacuum expected value of the Higgs
field. Examining $\epsilon_3$ does not help, but a plot (figure \ref{e6sym}) 
of the corresponding values of $\epsilon_6$ -the parameter that is
substituted by $\epsilon_3$- shows qualitative differences between
this scale and the others. In any case, this event forced upon us the need
to expand the search to other speculated particles.

\begin{figure}
    \centering
    \includegraphics[width=9cm]{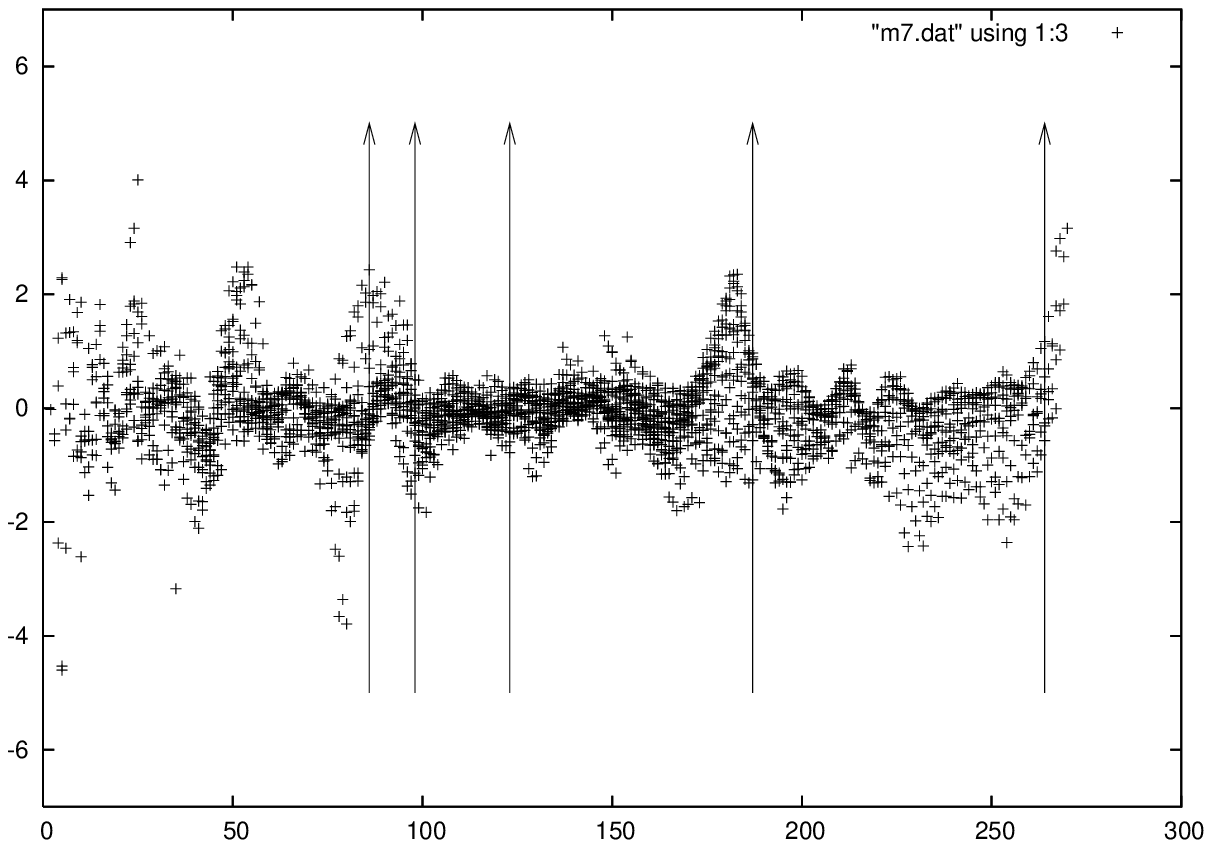}
    \caption{error in mass prediction for a model from  Takahiro Tachibana,
     Masahiro Uno, Masami Yamada, and So Yamada. Please refer to v2 for more examples}
     \label{m7}
\end{figure}

On a different take, if part of nuclear stability comes from unaccounted
interactions with
elementary particles, then an unexplored field is to look for deviations
in theoretical and semiempirical mass models. We took a look to this
in previous versions of this paper, which the reader could be interested
to glance. A typical example is figure \ref{m7}. Semiempirical models adjust
the parameters from some fit to determined nuclei, so a given model does not
need to show error in all the points we are looking to, and besides one
must consider other mathematical sources of deviation.

\begin{figure}
    \centering
  \includegraphics[width=9cm]{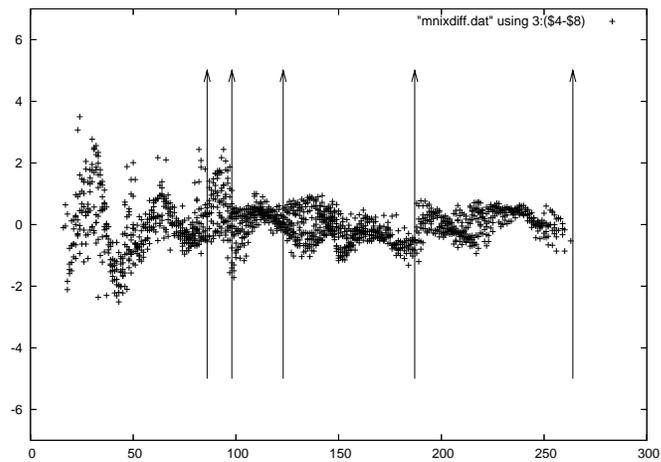}
    \caption{error in mass prediction for the FRDM model}
     \label{mnix}
\end{figure}

For comparison, we show in the figure \ref{mnix} the error plot from the FRDM, to 
confirm that it is excessively noisy in the low area and excessively corrected in 
the high. At these time we had not added the lower masses to our research.

\begin{figure}
\includegraphics[width=13.4cm]{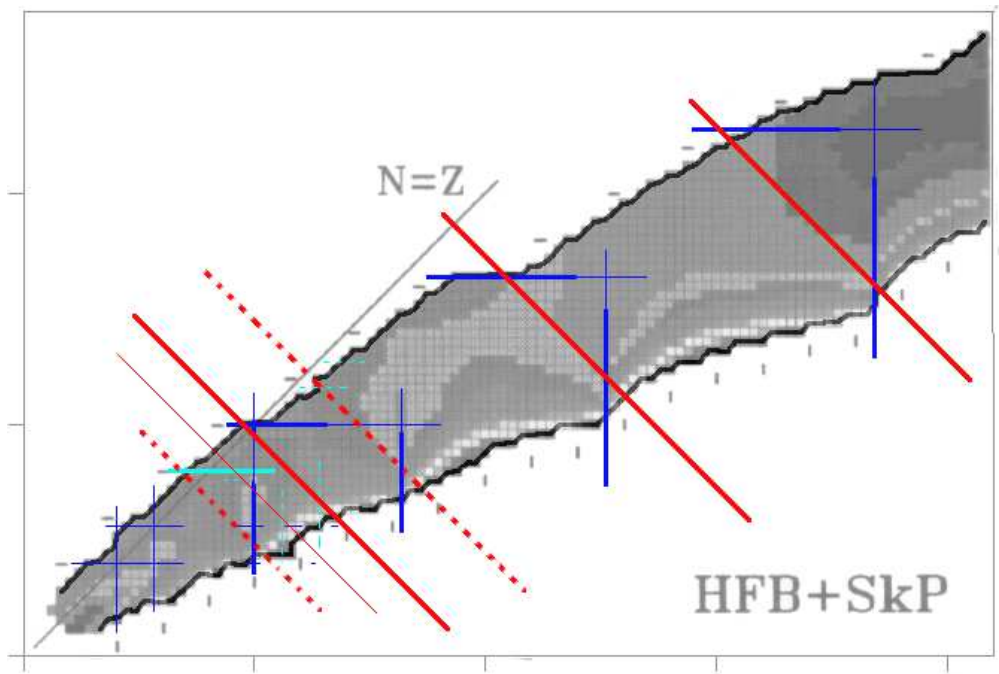}
\caption{Figure 7 of \cite{skins}, with our masses over-imposed.
 Diagonal isobars correspond
 to masses, horizontal and vertical lines to magic numbers. The background, taken
 from a (lost) HFB+SkP model,  shows the calculated difference between proton and neutron radius. 2N
 and 2P driplines are shown; check for instance \cite{uno} for 1N, 1P drip line predictions from other models }
\label{skp}
\end{figure}

Another interesting input comes from purely microscopical models with an empirical
 force. In figure \ref{skp} we see how close
the effects seem to follow our standard model particles. This model, from \cite{skins},
also calculates the 2-proton and 2-neutron driplines, and we can infer how the proton
magic numbers are generated at the proton dripline.

\section{HEP systematic error?}
Lacking still of a theoretical model, we are unable to say if this 
empirical analysis is a prediction of the Higgs or just a prediction
of an anomaly in the background when detecting energies in the 115 GeV
area. In the later case, the -nuclear- physics at the detectors would be 
the one to blame.

During experiments at CERN it has occurred sometimes that an excess of events has 
been announced for a very concrete energy range, but additional statistical analysis, or 
renewed experimental input, has smoothed down the initial deviations. Our
first three points, at
40 GeV \cite{UA1,nature}, 68 GeV \cite{L3a,L3b} and 115 GeV \cite{aleph}, are in
this category. 

We should to stop a moment to examine if all the three signals can be linked to an unlocated 
calibration problem. It could happen if an unforeseen non-uniformity in distributing 
energy causes some background events to accumulate, simulating to be a signal. If a 
naive nuclear mass model is used at some point in the physics calculations in detectors,
noise patterns similar to the ones in the previous section could propagate to the data.

Thus if a calibration or a detector depends, say, on secondary decay from near proton 
drip-line nuclei, a  mass model taking these masses into account should be used for the 
physics of the material. The same would happens with secondary decays near neutron 
drip line, but this line is rarely reached experimentally. If simple analytic 
approximations, such as Weisaker formula, are used instead, we can expect errors to 
happen related to these mass values. As all the CERN experiments share legacy code 
both for simulations and actual calibration of measurements, it could be that such 
approximation were hidden in some shared code.

Of the four values in CERN reach, three of them correspond very accurately to the 
troubled event excesses, while the other is just under the peak of the Z0 particle, so that 
a error there should be masked under the huge quantity of real Z0 events. 

On other hand, physical reality of the nuclear points could be claimed on 
grounds of the two extant masses, 175 and 246 
GeV, which do not correspond to CERN measures but are also commonly associated to 
high energy physics. 

Additional support for the physical reality of the excesses could be coming now from 
HERA \cite{carli} where some events have been reported about 40 GeV. A common bug 
between the code of HERA and CERN is more unlikely that between CERN 
experiments.

\section{A theoretical proposal}

Recoil corrections to the bound states of a relativistic  two body system
 $m<M$ can be expressed as a polynomial $P_m(1/M)$. If the small
 particle $m$ has an additional coupling to a yukawian field of mass
 M', one can use the shape of $P_{m,M'}(1/M)$ to determine the mass
 of the field $M'$. We call this method the {\it Lamb's Balance}. 
 
 The {\it Lamb's Balance Conjecture} is to suppose that the shape of $P_{m,M'}$
 presents a maximum when $M\sim M'$. The nuclear LBC assumes that
 this effect will be measurable at proton and neutron drip lines, where external
 nucleons are far from the rest of the nucleus ---neutrons even are
 in a distinctive skin--- and their exchanged momentum 
 is small, so that the rest of the nucleus can be seen as a single particle.
 
 In these conditions, the nuclear LBC would be able to assign a particle to each 
 magic number as they cross the drip lines.

 As before, the analysis is more
 or less model independent; one-nucleon or two nucleon driplines either 
 from FRDM95 or FRDM92 or from any other popular model can be used 
 without sensible differences (except for the neutron line around the values 
 of $W$-$Z0$). Besides, if $P_{m,M'}(1/M)$ decreases 
 smoothly after peaking, the effect will be noticed globally, for instance as 
 a contribution to the the surface terms of $M(A,Z)/A$ in liquid drop model.
 

 As an alternative to be researched too, the extra particles could influence directly
the spin orbit correction of the bound state, instead of its recoil correction. In any case
the influence would imply a missed subtlety in traditional scale separations as low momenta
electroweak interactions, where we always have felt justified to use Fermi constant as
an approximation to the Z0 and W propagators. It should happen that the existence
of a coupled virtual particle M' modifies the propagator in almost the same way
as if it were a real available mode, so the low energy limit should reflect it. The
author, presently, can not calculate if such unlikely modification does indeed exist.

\section{The role of charged bosons}

 Perhaps the more doubtful issue is the role of $W^\pm$, that,
 sandwiched between $Z0$ and the 68 GeV signal, only
 seems to correspond to weak semimagicities if any. But here one must remember
 the very special role that $W^\pm$ has in the nuclear table, as it is the responsible
 of the process of beta decay.  Then a virtual $W^-$, besides to contribute to the
 dressing of nucleons, is able to draw energy from them and to decay to electron plus 
 antineutrino; a process a lot more important that dressing and recoiling.

The point about any charged particle is that its role in the model above compites with
its role as a mediator of nuclear beta decay, when the nucleus has enough energy to
implement this decay path. A neutral boson, on the other side, does not have this
possibility for nuclear ground states, which are the ones we are studying. For a charged
scalar very weakly coupled to leptons, we can expect an equilibrated competition, while
for the W particle most of the eggs seem to be in the decay nest.

An histogram of all the known beta decays\footnote{See author's website,
at {\tt http://dftuz.unizar.es/\%7Erivero/research/}}, binned by atomic 
mass, shows a strong peak for
atomic mass of 81 GeV, and a previous step about 68 GeV, but unfortunately the strong
peak coincides with the extra stability from the magic nucleus Z=40, N=50. In fact the
strongest peak in the histogram comes from decays near other magic nucleus, Z=58, N=82.
Still, it could be possible to extract some information if the histogram is separated
according forbidness, parity, and change of angular momentum of the decay mode.

\section{Recapitulation and remarks}

We have extracted the following clues:
\begin{itemize}
\item A mass at 146 GeV, the scale of electroweak vacuum, contributes  the Z=114 and N=184 magic numbers.
\item A mass at 175 GeV, the one of top quark and its mesons, generates the Z=82 and N=126 magic
numbers
\item A mass at 115 GeV, the same signal that was detected at ALEPH, generates the N=82 magic
number. It should generate also the subshell closing around 60 protons, but the higgs coupling
to protons is expected to be lower, due to their different quark composition. 
\item A mass at 91 GeV, the one of the Z boson, generates the Z=50 magic number.  Perhaps
it helps to the N=50, Z=50 double magicity too, as well as to a subshell closing in N.
\item Extra noise is seen around the mass scale of the charged boson W; how this boson could
contribute to the N=50 magicity it is not clear, and probaly it doesn't.
\item A mass of 68 GeV, time ago suspected of a charged scalar, generates
the N=50 and Z=40 magic numbers.
\item A mass around 40-45 GeV, time ago examined as a candidate for the top quark,
 could contribute to the magicity of both N=28 and Z=28
\item Lower subshell values 20, 8, 2, would come mainly from the nuclear
well potential.
\end{itemize}

 If the signals are to be taken seriously, they
 strongly points towards a non minimal Higgs sector, say 2DHM, Zee, etc. When the LBC
 becomes a computational tool, it will be possible to distinguish a bit more between
 pseudoscalar, scalar, and vectorial particles.
 
 One particle should appear within a very few percent of the 264 GeV vacuum. This
 strongly restricts the parameters in model building. On other hand the mass at 
 175 GeV is known to be the one of the Top quark... If the Top couples to the nucleon
 after all, it should be by using its family of very short lived mesons, a very weak
 method. So another possibility is to suppose one of the extant bosons from the higgs sector to
 be mass-degenerated with the top quark. This degeneration between a boson
 and a fermion could be the only residual of a SUSY scheme.
 
In any case, the complete particle spectra surpasses the Minimal Standard Model, and it is not
compatible with expectations for the Minimal Supersymmetric Standard Model. 

Lets note
that recent research on Dimensional Deconstruction, by Georgi et al, has brought
a new family of models needing at least a pair of higgs doublets but independent
of the existence of supersymmetry. It could be worth to examine its compatibility
with our mass spectrum. Dimensional deconstruction gets symmetry breaking from extra
dimensions, a old theme whose pretties representative is the Connes Lott model. But
contrary to NCG, it does not requires a minimal higgs sector.

Respect to magicity, generation should be understood as contribution. This is 
because we have still the complementary,
traditional, contribution to spin orbit couplings from nuclear relativistic effects.
Still, these effects have never been able to justify completely the corrections
to binding energy. Traditionally it was considered that a lack of computing power
was the cause, and semi empirical models felt perfectly able to complement this
lack by fitting arbitrarily the spin orbit force. In view of the observed coincidences,
perhaps the arbitrariness should be constrained to depend on our masses.

Other explanations could be explored. For example, it could happen that the same
mathematical symmetry-breaking acts in nuclear physics and, for different 
causes, in elementary particle physics. Then the only remaining coincidence
would be the one between the end of the stability islands and the 
electroweak vacuum. Even if this is the case, it should be mathematically worth to
examine the mechanism in nuclear physics, because it includes both
the electroweak bosons and  top quark mass values in a same 
unifying schema.

The special role of external shells invites also to look for experimental methods such as
stop antiproton or antineutron bombarding in order to get more detail of the properties of this
kind of nucleus. Perhaps the final clue pointing to the Higgs is already buried in the data banks
of ancient CERN machines.

\section*{Appendix}

When referring to the Appendix in the previous version of this preprint,
please note this 
modification of point (6):
 Two quantities of 68 and 45 euro are to be added to the sequence.

 \end{document}